\documentclass[11pt]{amsart}
\usepackage[cp1251]{inputenc}

\usepackage{amssymb,amsmath,amscd,color}

\theoremstyle{plain}

 \newcommand{\cO}{\mathcal{O}}
\newcommand{\rr}{\mathbf{r}}

        \newcommand{\field}[1]{{\mathbb{#1}}}
        \newcommand{\NN}{\field{N}}
        \newcommand{\ZZ}{\field{Z}}
        
        \newcommand{\RR}{\field{R}}
        \newcommand{\CC}{\field{C}}

\allowdisplaybreaks

\begin{document}

\title[]{Geometry and quasi-classical quantization of magnetic monopoles}

\author[]{I.A. Taimanov}
\address{Sobolev Institute of Mathematics of SB RAS, 630090 Novosibirsk, Russia}\email{taimanov@math.nsc.ru}

\thanks{}

\begin{abstract}
We present the basic physical and mathematical ideas (P. Curie, Darboux, Poincar\'e, Dirac) that led to the concept of magnetic charge, the general construction of magnetic Laplacians for magnetic monopoles on Riemannian manifolds, and the results of Yu.A. Kordyukov and the author on the quasi-classical approximation for the eigensections of these operators.
\end{abstract}

\date{}

 \maketitle
\section{Introduction}

This article concerns the scheme of quasi-classical quantization of magnetic monopoles proposed by Yu.A. Kordyukov and the author in the article \cite{KT20} which is related to the articles of the same authors on the trace formulas for magnetic Laplacians \cite{KT19,KT21}. It is an expanded presentation of the author’s report at the international conference on mathematical physics, dedicated to the centenary of the birth of V. S. Vladimirov in January 2023 at the Steklov Mathematical Institute of RAS.

We consider in detail the background of the issue.

In \S 2 the works of P. Curie, Darboux and Poincar\'e are discussed, in which magnetic monopoles in classical electrodynamics were considered from various points of view.

In \S 3 there is exposed the derivation of the condition for the quantization of electric charges, for the sake of which Dirac introduced magnetic monopoles in quantum mechanics. This idea turned out to be fruitful; the current state of its development is presented in the monograph \cite{Shnir}.

The general construction of magnetic Laplacians is presented in \S 4.
   
In \S\S 5-6 we present at the physical level the construc\-tion of quasi-classical quantization of the magnetic Laplacian. It is based on an extension of the multidimensional
the WKB method (Maslov’s canonical operator) for the case of operators acting on sections of nontrivial bundles. This presentation makes the construction transparent, and for rigorous mathe\-matical details we refer to \cite{KT20}.

\section{Curie, Poincar\'e and magnetic monopole}

Apparently, for the first time from a physical point of view, magnetic monopoles
were discussed by Curie. In a short note \cite{Curie} he cited some physical
consequences of their existence in 1894. Curie tried to use these conclusions
in his experiments to detect monopoles, which did not lead to success. However, he pointed out that

{\sl ``from the point of view of energy, from the point of view of symmetry,
one can imagine without absurdity the currents of magnetism and
free magnetic charges. Of course, it would be rash to draw conclusions from this,
that these phenomena really exist. But if that were the case,
they must satisfy the conditions we have specified.''}
\footnote{``...au point de vue de l'\'energ\'etique, au point de vue de la sym\'etrie,
on peut concevoir sans absurdit\'e les courants de magn\'etisme et les
charges de magn\'etisme libre. Il serait certes t\'em\'eraire d’induire de l\`a
que ces ph\'enom`enes existent r\'eellement. Si cependant il en \'etait ainsi,
ils devraient satisfaire aux conditions que nous avons \'enonc\'ees.''}

In 1896, Poincar\'e made an attempt to explain the results of experiments by Birkeland, a Norwegian physicist and his former student, with Crookes tubes (electric discharge tubes). The cathode rays discovered in them were then the subject of intensive research.
Poincar\'e proposed a mathematical model of the behavior of cathode rays discovered by Birkeland \cite{Poincare}.

He started with the following assumption:

{\sl ``Let's write the equations of the cathode ray, likening it to a rapidly moving material electrically charged particle''}
\footnote{``...nous ecrirons les equations du rayon cathodique, en l'assimilant a une particule materielle en mouvement rapide, chargee d'electricite...''}

and onwards

{\sl ``Assume that there is a single magnetic pole, which we will take as the origin of coordinates''}
\footnote{``Supposons un seul pole magnetique, que nous prendrons pour l'origine...''.}
.

After this, Poincar\'e wrote out the equations of motion $\rr(t) = (x(t)$, $y(t)$, $z(t))$ of a charged particle in a magnetic field with a singularity at the origin:
$$
\frac{d^2 x}{dt^2} = \frac{\lambda}{r^3}\left(y \frac{dz}{dt} - z\frac{dy}{dt}\right) ,
$$
\begin{equation}
\label{poincare}
\frac{d^2 y}{dt^2} = \frac{\lambda}{r^3}\left(z \frac{dx}{dt} - x\frac{dz}{dt}\right) ,
\end{equation}
$$
\frac{d^2 z}{dt^2} = \frac{\lambda}{r^3}\left(x \frac{dy}{dt} - y\frac{dx}{dt}\right) ,
$$
where $r = |\rr|$ and $\lambda$ is a constant depending on the intensity of the magnetic field and the ``nature of the cathode rays''
\footnote{``c'est-a-dire, dans l'hypothese de Crookes, de la masse de la particule materielle en mouvement et de sa charge electrique'' --- {\it transl.}:\ ``that is, according to Crookes' hypothesis, from the mass of a moving material particle and its electric charge''.}
.

Denoting by $\dot{\rr}$ and $\ddot{\rr}$ the speed and the acceleration of the particle, we rewrite these equations in the form
$$
\ddot{\rr} = \frac{\lambda}{r^3} [\rr \times \dot{\rr}].
$$
From \eqref{poincare} Poincar\'e derives the law of conservation of energy  
$$
|\dot{\rr}|^2 = C = \mathrm{const}
$$
and an important relationship
$$
r^2 = Ct^2 + 2Bt + A,
$$
Really,
$$
\frac{d^2 r^2}{dt^2} = \frac{d^2 \langle \rr, \rr \rangle}{dt^2} =
2 \langle \dot{\rr}, \dot{\rr} \rangle + 2 \langle \rr, \ddot{\rr}\rangle = 2 |\dot{\rr}|^2 = 2C,
$$
where by $\langle \cdot,\cdot\rangle$ we denote here and below the scalar (Euclidean or Hermitian) product of vectors.
A direct consequence of the basic equations is also that the vector $[\rr \times \dot{\rr}] + \frac{\lambda}{r}\rr$ is constant and we get three more conservation laws:
\begin{equation}
\label{conev}
[\rr \times \dot{\rr}] +\frac{\lambda}{r} \rr = v=
\begin{pmatrix}
a \\ b \\ c \end{pmatrix},
\end{equation}
where the vector $v$ is constant along the trajectories.
It immediately follows that
\begin{equation}
\label{cone}
\langle v,\rr \rangle = ax+by+cz = \lambda r,
\end{equation}
i.e., the particle moves along a cone of rotation with its vertex at the origin.
Since the acceleration vector $\ddot{\rr}$ at every point of the cone, except its vertex, is perpendicular to the velocity vector $\dot{\rr}$ and the generator $\rr$ of this cone, then
the trajectory is geodesic.

Thus, the trajectories of a particle of mass $m$ and charge $q$ in a magnetic field (monopole)
$$
{\bf H} = \frac{H_0}{r^3}\rr, \ \ \ H_0 = -\frac{m c \lambda}q,
$$
where $c$ is the speed of light in vacuum,
are geodesics on the cones \eqref{cone}, where the vectors $v$ are determined by the initial data according to \eqref{conev}.

Note that this Poincar\'e explanation depends on Crookes's assumption that cathode lines are streams of massive charged particles
\footnote{``si l'hypothese de Crookes n'est pas vraie, il semble bien que tout se passe comme se elle l'etait''
--- {\it transl..}: ``if Crookes' hypothesis is not true, it seems that everything happens as if it were''.}
.

Moreover, he emphasizes that ``the theory is incomplete because we assume the existence of a single magnetic charge''
\footnote{``Est-ce que la theorie est incomplete, parce que nous avons suppose un pole magnetique unique''.}
.

The equation \eqref{poincare}, as it turned out, had already been considered in 1878 by Darboux in a short note
\cite{Darboux}, which did not assume the existence of a magnetic monopole, but was also related to the theory of magnetism. Darboux solved the problem of finding
equilibrium of a flexible and inextensible weightless wire conducting current and
under the influence of a magnet pole
\footnote{Annotation to \cite{Darboux}:
``Trouver la figure d'\'equilibre d'un fil flexible et inextensible non pesant, travers\'e par un courant et sounis \`a l'influence du p\'ole d'un aimant''.}
.
Darboux reduced it to the equation \eqref{poincare} and described the resulting curves
as geodesics on cones of rotation, completely analogous to what Poincar\'e did later (he apparently was not familiar with Darboux’s work and did not refer to it).

A year after Poincar\'e's work, while studying cathode rays, Thomson showed that they were streams of charged particles, which were called electrons
\cite{Thomson}.

\section{The Dirac monopole}

In Dirac's work \cite{Dirac31}, which occupies just over 12 pages, the first page is devoted to the development of the mathematical foundations of modern physics and the reasoning given on it has still not lost its relevance, the second page contains a discussion of the existence of strange particles that have the same mass as and electron, but negative energy. It was noted that they can be interpreted as holes in the unobservable fully filled distribution of negative energy states. As Dirac notes,

{\sl ``A hole, if there were one, would be a new kind of particle, unknown to experimental physics, having the same mass and opposite charge to an electron. We may call such a particle an anti-electron. We should not expect to find any of them in nature, on account of their rapid rate of recombination with electrons, but if they could be produced experimentally in high vacuum they would ne quite stable and amenable to observation''.}

The existence of such particles followed from the famous Dirac equation, derived in 1928.
A year after the publication of \cite{Dirac31}, in 1932, these particles were discovered by Anderson and are now known as positrons.

Next, Dirac moves on to the main goal of the article, which is

{\sl ``to  put  forward  a  new  idea  which  is  in many  respects  comparable  with  this  one  about  negative  energies. It  will be concerned  essentially,  not with electrons  and protons,  but with the reason for the  existence  of a  smallest  electric  charge''.}

Let the wave function
$\psi(x,y,z,t)$ describe the motion of the particle. Let us present it in the form
$$
\psi = A e^{i \gamma},
$$
where $A$ and $\gamma$ are real-valued functions.
Assuming $\psi$ normalized, it determines the state up to an arbitrary constant factor which is equal in the
absolute value to zero. This means that we can assume that $\gamma$ does not have a specific value at a point and for two points the phase difference is determined with respect to the curve connecting them.

For a pair of wave functions $\varphi$ and $\psi$ the absolute value of the quantity
$$
\langle \varphi | \psi \rangle = \int \bar{\varphi}\psi\, dx \,dy \, dz
$$
has a physical sense and, as Dirac notes, we must assume that

{\sl ``The change in phase  of a  wave function round any closed curve must  be the  same for  all  the  wave functions''.}

These requirements and the superposition principle can be achieved as follows.
Let
$$
\psi = \psi_1 e^{i \beta},
$$
where $\psi_1$ is a wave function with a phase defined at each point, and the phase uncertainty of
$\psi$ is expressed as a factor $e^{i\beta}$.
Let us assume that the phase is not defined at each point, but its derivatives
$$
\varkappa_x = \frac{\partial \beta}{\partial x}, \ \
\varkappa_y = \frac{\partial \beta}{\partial y}, \ \
\varkappa_z = \frac{\partial \beta}{\partial z}, \ \
\varkappa_t = \frac{\partial \beta}{\partial t}
$$
are defined, but are not required to satisfy the integrability conditions
$$
\frac{\partial \varkappa_x}{\partial y} = \frac{\partial \varkappa_y}{\partial x} \ \ \dots \ \ \mbox{etc.}
$$
If $\psi$ satisfies some equation containing the momentum and the energy operators ${\bf p}$ and $W$, then 
$\psi_1$ satisfies the same equation in which ${\bf p}$ and $W $ are replaced with
${\bf p}+ h \varkappa$ and $W - h \varkappa_0$. Thus, the transition from $\psi$ to $\psi_1$ is reduced to the inclusion of an electromagnetic field with the potential
$$
A = \frac{\hbar c}{e} \varkappa,
$$
where $h$ is Planck's constant, $c$ is the speed of light in vacuum, and $e$ is the elementary charge (a positive charge which equal in the absolute value to the charge of an electron). At the same time, Dirac notes that he is considering a particle with a charge $q=-e$. Thus the general formula for $A$
takes the form
\begin{equation}
\label{charge}
A = -\frac{\hbar c}{q} \varkappa,
\end{equation}
where $q$ is the charge of the particle for which $\psi$ is the wave function.

Dirac notes that the connection between the phase nonintegrability and the electromagnetic field is a manifestation of the Weyl principle of gauge invariance.
However, there are two additional circumstances:

1) the phase is always determined up to a multiple of $2\pi$;

2) changes in the phases of various wave functions along closed curves can be different and differ by values that are multiples of $2\pi$.

If 1) is obvious, then let us focus on manifestations of 2). The phase change along a closed curve
$\tau$ in $x,y,z$-space, according to Stokes’ theorem, is equal to
$$
\int_\tau (\varkappa_x dx + \varkappa_y dy + \varkappa_z dz) = \int_\Gamma (\mathrm{curl}\, \varkappa, dS),
$$
where $dS$ is the surface area element of $\Gamma$ bounded 
by $\tau$. If at some point $P$ we have $\psi(P)\neq 0$, then when contracting closed curves that go around this point, the phase change tends to zero. If, $\psi(P)=0$, then
we can only say that
$$
\frac{e}{\hbar c} \int_\Gamma ({\bf H},dS) = 2\pi n_\tau,
$$
where ${\bf H} = \frac{\hbar c}{e} \mathrm{curl}\, \varkappa$ is the magnetic field and $\tau = \partial \Gamma$.

If we have a closed surface $\Gamma$ such that it is divided into small domains containing one zero of the function $\psi$, we, adding the previous relation over all such regions, obtain
\begin{equation}
\label{quantum}
\frac{e}{\hbar c} \int_\Gamma ({\bf H},dS) = 2\pi n, \ \ n \in \ZZ,
\end{equation}
i.e., the magnetic flux of such a system is quantized and is a multiple of $2\pi$.
For $N \neq 0$, the domain bounded by the surface $\Gamma$ must contain magnetic charges.

Let us consider the simplest case of a single-point magnetic charge located at the origin of coordinates and assume that the electric field is zero.
Then the magnetic field has the form
\begin{equation}
\label{mono}
{\bf H} = g_D \frac{\bf r}{r^3},
\end{equation}
where ${\bf r}$ is the position vector of the point and $r = |{\bf r}|$.
Let us consider the spherical coordinates $r,\varphi,\theta$.
As the vector potential of this magnetic field we can take
$$
\varkappa_\theta = \varkappa_r=0, \ \ \ \varkappa_\varphi = \frac{1}{2r}\tan \frac{\theta}{2}.
$$
In this case, using separation of variables, we represent $\psi_1$ in the form
$$
\psi_1 = f(r) S(\theta,\varphi),
$$
the wave equation
$$
-\frac{\hbar ^2}{2m}\nabla^2 \psi = W\psi
$$
for $\psi = \psi_1 e^{i\beta}$
splits into the following system
$$
\left\{\frac{d^2}{dr^2} + \frac{2}{r} \frac{d}{dr} - \frac{E}{r^2}\right\} = - \frac{2mW}{\hbar^2}f,
$$
\begin{equation}
\label{tamm2}
-\left\{\frac{1}{\sin \theta} \frac{\partial}{\partial \theta} \sin \theta \frac{\partial}{\partial \theta} +
\frac{1}{\sin^2 \theta}\frac{\partial^2}{\partial \varphi^2} +
\frac{i}{2}\sec^2\frac{\theta}{2}\frac{\partial}{\partial \varphi} -\frac{1}{4}\tan^2\frac{\theta}{2}\right\} S =
E S.
\end{equation}
We rewrite the second of these equations as
$$
\Delta^L S = E S
$$
and call the operator $\Delta^L$ the magnetic Laplacian on the two-dimensional unit sphere.
We will explain the general definition of  this operator and the meaning of the symbol $L$ in the next section.

In this case, the general solution to the wave equation was obtained by Tamm \cite{Tamm}
\footnote{Dirac refers to this work, published a month after \cite{Dirac31}.}.
He showed that the eigenvalues of $E_N$, $N=0,1,\dots$, have the form
$$
E_N = N^2 + 2N + \frac{1}{2}
$$
and multiplicities $2N+2$. A basis of eigenfunctions at the smallest eigenvalue
$E = \frac{1}{2}$ is given by ``functions''
\begin{equation}
\label{tamm}
S_a = \cos \frac{\theta}{2}, \ \ S_b = \sin \frac{\theta}{2} e^{-i\varphi}.
\end{equation}
Moreover, $S_a$ is continuous everywhere, and $S_b$ has a singularity at $\theta = \pi$ and its phase changes by $2\pi$ when going around this point along a small contour.

The basis \eqref{tamm} was obtained by Tamm \cite{Tamm}.
In Dirac's work \cite{Dirac31} it was reproduced with a typo: there was an incorrect sign for $\varphi$ in the definition of $S_b$. The formula \eqref{b} given below shows that $S_a$ and $S_b$ as sections of a line bundle with $c_1=-1$ have no singularities.

However, Dirac's main goal was an important physical conclusion. From the quantization condition \eqref{quantum}
follows that
\begin{equation}
\label{diracq}
e g_D = \frac{\hbar c}{2} n, \ \ n\in \ZZ.
\end{equation}
In \cite[\S1 ]{Dirac31} $e$ denotes the ``smallest'' electric charge,
but in \cite[\S 3]{Dirac31} the charge of an arbitrary particle is taken as $e$. Therefore \eqref{diracq} is {\it the condition for quantizing the charges of electric particles}.
As Dirac noted,
{\sl ``. . . if there exists any monopole at all in the universe, all
electric charges would have to be such that $e$ times this monopole strength is equal
to $\frac{1}{2} n \hbar c$''}
\cite{Dirac78}.

The Dirac monopole has not yet been discovered.
Let us evaluate its physical characteristics.

Since in the CGS system the fine structure constant has the form
$$
\alpha = \frac{e^2}{\hbar c} \approx \frac{1}{137},
$$
we rewrite the quantization condition \eqref{quantum} for $n=1$ in the form
$$
g_D \approx \frac{137}{2} e.
$$

We cannot estimate the mass of the monopole (as well as the electron) from any theoretical assumptions.
But if, for example, we assume that the classical radii of the electron and magnetic monopole coincide:
$$
r_e = r_D,
$$
where
$$
r_e = \frac{e^2}{4\pi \varepsilon_0 m_e c^2}, \ \
r_D = \frac{g_D^2}{4\pi \mu_0 m_D c^2}
$$
and $\varepsilon_0$ and $\mu_0$ are the electric and the magnetic constants, then
$$
m_D = \frac{g_D^2}{e^2} m_e = \frac{1}{(2\alpha)^2} m_e \approx 4692 m_e \approx 2.4 \,\mbox{GeV}.
$$

\section{Magnetic Laplacian}

We cited Dirac's reasoning above to show how he approached from a physical point of view such an important concept as a $U(1)$ bundle and the condition for quantizing the Chern classes of such bundles. An explanation of Dirac's work via these topological concepts
was given in \cite{WY75,WY}.

We will consider a more general situation where the system is defined on
a $d$-dimensional differentiable manifold $M$.We assume that all objects on $M$ which are forms,
connectivity, metrics, etc., are differentiable the required number of times.

By a magnetic field we mean the closed $2$-form $F^{(0)}$:
$$
F^{(0)} = \sum F^{(0)}_{jk} dx^j \wedge dx^k, \ \ dF^{(0)} = 0.
$$
The vector potential of this form is defined as
\begin{equation}
\label{curvature}
A^{(0)} = \left(A^{(0)}_k\right), \ \ \ \frac{\partial A^{(0)}_k}{\partial x^j} - \frac {\partial A^{(0)}_j}{\partial x^k} =
F^{(0)}_{jk}.
\end{equation}
We assume that $c=\hbar=1$, which is accepted in
the natural units system.

If the form $F^{(0)}$ is not cohomologous to zero, then the vector potential is determined only locally,
in areas to which $F^{(0)}$ is restricted to an exact form.

Let $M$ be endowed with a Riemannian metric
$$
g_{jk} dx^j dx^k
$$
(hereinafter we mean summation over upper and lower repeating indices), which is used to determine the Hamiltonian (kinetic energy) of particle motion on $M$:
$$
H = \frac{m}{2} g^{jk} p_j p_k,
$$
where $g^{jk}$ is the inverse tensor to $g_{jk}: g^{jk}g_{kl} = \delta^j_l$, and ${\bf p} =(p_1,\dots,p_d )^\top$ is the momentum of a particle. Next, for simplicity, we set $m=1$.
Let us rewrite the Hamiltonian in the form
$$
2H = \frac{1}{\sqrt{g}} p_j \sqrt{g} g^{jk} p_k, \ \ \ g = \det\left(g_{jk}\right)
$$
and replace the momentum with the momentum operator corresponding to it by quantization
$$
p_j \to -i \frac{\partial}{\partial x^j},
$$
having obtained the Laplace--Beltrami operator as a result of quantizing the Hamiltonian of a free particle:
\begin{equation}
\label{lb}
\Delta = -\frac{1}{\sqrt{g}}\frac{\partial}{\partial x^j} \sqrt{g} g^{jk} \frac{\partial}{\partial x^ k}.
\end{equation}
This operator is given by a covariant expression, i.e. is invariant under changes of coordinates.

Let us now assume that a particle has a charge
$$
q=Ze.
$$
In this case, the inclusion of the magnetic field $F^{(0)}$ interacting with the particle consists of
replacing the momentum with
$$
p_j \to p_j - q A^{(0)}_j.
$$
For brevity, we introduce the notation
$$
F = q F^{(0)}, \ \ A = q A^{(0)}, \ \ dA = F.
$$
The quantization results in the replacement of the Laplace--Beltrami operator with the magnetic Laplacian:
\begin{equation}
\label{magnetic}
\Delta = -\frac{1}{\sqrt{g}}\left(\frac{\partial}{\partial x^j} - iA_j\right) \sqrt{g} g^{jk}
\left(\frac{\partial}{\partial x^k} - iA_k\right).
\end{equation}

Dirac's arguments about the ambiguity of functions $\psi$ are reformulated as follows:

\begin{itemize}
\item
$\psi$ are sections of a complex line (one-dimensional) bundle $L$ with structure group
$U(1)$ (from now on we will simply talk about $U(1)$-bundles);

\item
the vector potential $A = \left(A_k \right)$ defines a connection
$$
\nabla_k = \frac{\partial}{\partial x^k} - iA_k
$$
on $L$;

\item
on sections $\psi$ of the bundle $L$ there act (locally) gauge transformations:
$$
\psi(x) \to \eta(x) \cdot \psi(x), \ \ \eta(x) = e^{i f(x)} \in U(1);
$$

\item
since the requirement of connection invariance under the gauge trans\-for\-mation must be satisfied:
$$
\psi \to \tilde{\psi} = \eta\cdot \psi, \ \ \ \widetilde{\nabla}_k \tilde{\psi} =
\eta\cdot \left(\nabla_k \psi\right), \ \ k=1,\dots,d,
$$
then the connection is transformed according to the formula
\begin{equation}
\label{gaugecon}
A_k(x) \to \tilde{A}_k(x) = \eta(x) \cdot A_k(x) \cdot \eta^{-1}(x) + i \frac{\partial \eta(x )}{\partial x^k}\eta^{-1}(x) = A_k - \frac{\partial f}{\partial x^k};
\end{equation}

\item
by \eqref{gaugecon}, $F$ and the difference $A - A^\prime$ of any two connections on the same bundle are correctly defined $2$- and $1$-forms on $M$;
\item
by \eqref{curvature}, $F$ is the curvature form of the connection $A$;

\item
{\sl the quantization condition}: the curvature form $F$ is closed and after division by $2\pi$
defines an integer cohomology class:
\begin{equation}
\label{chern}
c_1(L) = \left[\frac{F}{2\pi}\right] \in H^2 (M;\ZZ),
\end{equation}
which is the first Chern class of the bundle $L$.
\end{itemize}

Note that if the quantization condition is not satisfied, then $A$ does not define a connection in
any line bundle.

The magnetic Laplacian
$$
\Delta = - \frac{1}{\sqrt{g}}\nabla_j \sqrt{g}g^{jk}\nabla_k
$$
is invariant under gauge transformations:
$$
\widetilde{\Delta} (\eta\cdot \psi) = \eta\cdot \Delta\psi,
$$
but its definition depends significantly on the choice of
a connection $A$ with the given curvature form (``magnetic field'') $F$.

Let us assume that connections $A$ and $A^\prime$ define the operators $\Delta$
and $\Delta^\prime$, corresponding to the same magnetic field $F$ and acting on sections of the same bundle $L$. Then their difference
$Q = A^\prime-A$ is a closed $1$-form: $dQ = q(F - F)=0$.

If the closed form $Q = A^\prime-A$ realizes the zero class of one-dimensional cohomology:
$$
[A^\prime-A] = 0 \in H^1(M;\RR),
$$
then $A^\prime-A = df$, where $f:M \to \RR$ is some function.
Then, by \eqref{gaugecon}, from
$$
\Delta \psi = E \psi
$$
follows that
$$
\Delta^\prime (e^{-if}\psi) = E (e^{-if}\psi),
$$
and thus the spectrum of the magnetic Laplacian does not depend on the choice of connection $qA$.

If the bundle $L$ is trivial, then, by choosing its trivialization $L = M \times \CC$, one can define a connection on it using the $1$-form $A = A_k dx^k$ on the manifold $M$.

\vskip3mm

{\sc Example. ``Dirac's monopole'' on a two-dimensional sphere.}.

Let us consider a two-dimensional unit sphere $M = S^2$ in the three-dimensional space and restrict to it
the $2$-form corresponding to the magnetic field \eqref{mono}. We obtain
$$
F^{(0)} = g_D \sin \theta\, d\theta \wedge d\varphi.
$$
The quantization condition gives
$$
2eg_D \in \ZZ.
$$
In the domain obtained from the sphere by puncturing the lower pole ($\theta = \pi$), the vector potential
of magnetic field $F^{(0)}$ can be taken in the form
$$
A^{(0)}_\theta = 0, \ \ A^{(0)}_\varphi = g_D (1-\cos \theta).
$$
Let us multiply $A^{(0)}$ by the charge of the particle
$$
q= Ze, \ \ \ Z \in \ZZ.
$$
For
$$
eg_D = \frac{1}{2}, \ \ \ q=Ze, \ \ \ Z=-1 \ \mbox{(electron)},
$$
by substituting the expression for $A = qA^{(0)}$ into \eqref{magnetic}, we obtain, exactly,
operator $\Delta$ from equation \eqref{tamm2}.

By \cite{Tamm}, the smallest eigenvalue for this operator is
$\lambda = \frac{1}{2}$. The basis of eigensections can be chosen in the form
$$
S_a = \cos \frac{\theta}{2}, S_b = \sin \frac{\theta}{2} e^{-i\varphi}.
$$
We have already mentioned Dirac's remark that $S_b$ has a singularity at
$\theta = 0$. As Wu and Yang first showed, these ``functions'' are sections of the line bundle $L$ and  have no singularities \cite{WY}.

$U(1)$-bundles over a two-dimensional sphere are classified by the first Chern class. Let us consider the stereographic projection of the unit sphere $x^2+y^2+z^2=1$ onto the plane $x,y$, which we will identify with the complex line with the coordinate $z=x+iy$. This projection preserves orientations and identifies the two-dimensional sphere with the complex projective line
$\CC P^1 = \CC \cup \{\infty\}$. The complex line bundle $L$ with $c_1(L)=n$ over $\CC P^1$ is denoted by $\cO(n)$.

Any $U(1)$-bundle, i.e. a complex one-dimensional bundle with structure group
$U(1)$, over $\CC P^1$ is determined by the clutching function.

Let $[u:v]$ be homogeneous coordinates on $\CC P^1$: a point on
$\CC P^1$ is given by a pair of coordinates $[u:v]$, which are not both equal to zero and
are determined up to multiplication by a non-zero constant, i.e.
$[u:v] \sim [\lambda u : \lambda v], \lambda \neq 0$.
The sphere is represented as a union of overlapping domains
$$
U_a = \{ u \neq 0\}, \ \ U_b = \{ v \neq 0\}
$$
or like gluing of two disks
$$
D_a = \{ [1:z] \in U_a, |z| \leq 1\}, \ \ D_b= \{ [w:1] \in U_b, |w| \leq 1\}
$$
along the common boundary.
Since the disks $D_a$ and $D_b$ are contractible, the restrictions of any line bundle on them are trivial. Two trivial bundles $D_a \times \CC$ and $D_b \times \CC$ are glued together along
the boundaries into a line bundle over $\CC P^1$. The gluing is determined by the clutching function $\gamma(z),
z = e^{i\varphi}, |\gamma(z)| =1$, according to the formula
$$
(w = z^{-1},\mu) = (w, \gamma(z) \lambda) \sim (z, \lambda).
$$
A priory
$$
\gamma: \{|z|=1\} = S^1 \to U(1) \approx S^1
$$
and the topological type of the bundle is determined by the homotopy class of the mapping
$\gamma: S^1 \to S^1$. Any mapping from $S^1$ to $S^1$ is homotopic to a mapping of the form
$\varphi \to e^{ik\varphi}$, where $k \in \ZZ$.

Let us take as an example the tautological bundle $\cO(-1)$.
Since the point $[u:v] \in \CC P^1$ corresponds to a line with a direction vector $(u,v)$ in $\CC^2$, all such lines form a line bundle over $\CC P^ 1$. On $D_a$ the bundle is trivialized
$D_a \times \CC$ and in this case the pair of coordinates $(z,\lambda)$ corresponds to a point with coordinates $(\lambda,\lambda z)$. Similarly, when $D_b \times \CC$ is trivialized, the correspondence $(w,\mu) \to (\mu w,\mu)$ holds. These bundles are glued along the boundaries $D_a$ and $D_b$ according to obvious rules
$$
(\lambda, \lambda z) = \left(\frac{\mu}{z}, \mu\right) \ \ \mbox{at $\mu = z \lambda$}.
$$
Therefore, for $\cO(-1)$ the clutching function has the form
$$
\gamma(\varphi) = e^{i\varphi}.
$$
For the dual bundle $\cO(1) = \cO(-1)^\ast$ the clutching function is $e^{-i\varphi}$.
For the tensor powers of these bundles $\cO(-1)^k$ and $\cO(1)^k$, $k>0$,
the clutching functions are $e^{-ik\varphi}$ and $e^{ik\varphi}$. Obviously, for a trivial bundle $\CC P^1 \times \CC$ for
the clutching function can be set to $\gamma = 1$. Since
$$
c_1(\cO(-1)) = -1
$$
and $c_1(L \otimes L^\prime) = c_1(L) + c_1(L^\prime)$ for any pair of line bundles
$L$ and $L^\prime$, 
$c_1(\cO(k)) = k$ and for any line bundle $L$ over $\CC P^1$
the clutching function can be taken in the form
$$
\gamma_L = e^{-ic_1(L)\varphi}.
$$

For the Dirac monopole with $qg_D = - \frac{1}{2}$ we have
$$
c_1 = 2qg_D= -1.
$$
In $U_a$ a basis of eigensections corresponding to the eigenvalue $E=1/2$ is given by the sections
$$
S_a = \cos \frac{\theta}{2}, \ \ S_b = \sin \frac{\theta}{2} e^{-i\varphi},
$$
which in $U_b$, by the form of $\gamma_L$, are as follows
\begin{equation}
\label{b}
S_a = \cos \frac{\theta}{2} e^{i\varphi}, \ \ S_b = \sin \frac{\theta}{2}
\end{equation}
and have no singularities. They are sections of the tautological bundle $\cO(-1)$ over $\CC P^1$.

Explicit formulas for monopole harmonics, which are the eigensections of magnetic Laplacians corresponding to constant magnetic fields on a two-dimensional sphere, are derived in \cite{WY}.

Obviously, if $\Delta^L \psi = E \psi$, then, by complexly conjugating the analytical
expression for $\psi$, we obtain the section $\bar{\psi}$ of the conjugate bundle $L^\ast = \bar{L}$
and $\Delta^{L^\ast}\bar{\psi} = E\bar{\psi}$. Physically this is interpreted as an inversion of the charge of particle.

\section{Quasi-classical quantization of the Schr\"odinger operator}

In quantum mechanics, the WKB method allows one to construct (quasi-classical) approximations of the eigenfunctions of wave operators. Since in this case ``Planck's constant'' $\hbar$ is a small parameter, the powers of which are used to construct approximations, in what follows we will not switch to a natural units system and assume it is equal to unity.

Let us recall the WKB method for the one-dimensional Schr\"odinger equation:
$$
\left(-\frac{\hbar^2}{2m} \frac{d^2}{dx^2} + U(x) \right) \psi = E \psi.
$$
We will look for a solution in the form
$$
\psi(x) = e^{\frac{i}{\hbar} \sum_{k\geq 0} f_k(x) \left(\frac{\hbar}{i}\right)^k}.
$$
Let us substitute this approximation into the Schr\"odinger equation and formally expand the resulting expression in powers of $\hbar$. After dividing by $\psi$ we obtain
$$
\left[\frac{1}{2m}f_0^{\prime 2} + U - E\right] +
\left[ -\frac{1}{2m} (f_0^{\prime\prime} + 2 f_0^\prime f_1^\prime)\right] \hbar =
0\ \mathrm{mod}\, O(\hbar^2).
$$
Equating the free term on the left-hand side to zero, we deduce that
$$
f_0^\prime = \pm \sqrt{2m(E-U(x))}.
$$
Since the total energy of a particle of mass $m$ moving in a straight line in a potential field
$U(x)$, is equal to
$\frac{p^2}{2m} + U(x) = E$,
where $p$ is the momentum of the particle, then
$$
f_0^\prime = \pm \sqrt{p^2} = \pm p,
$$
and, by this formula,
$$
f_1 = -\frac{1}{2} \log p.
$$
We obtain a quasi-classical eigenfunction in the form
\begin{equation}
\label{quasi1}
\psi = \frac{C_1}{\sqrt{p}} e^{\frac{i}{\hbar} \int pdx} + \frac{C_2}{\sqrt{p}} e^{-\frac {i}{\hbar}\int pdx},
\end{equation}
where $C_1$ and $C_2$ are constants.

If we have the motion of a point in a potential well $\{ U(x) \leq E\}$ bounded by
 a pair of points at which $U(x)=E$, then outside this well the momentum $p$ becomes purely imaginary.
By choosing constants $C_1$ and $C_2$,  we would like to achieve the following situation 
 that outside the well this formula gives exponentially decaying solutions. This can be done if 
 the Bohr--Sommerfeld condition is satisfied
  \begin{equation}
  \label{bz}
  \frac{1}{2\pi \hbar} \oint pdx = n + \frac{1}{2}, \ \ n \in \ZZ,
  \end{equation}
 where the integral is taken over the closed trajectory of a particle in a potential well. Naturally, this is not possible for all values of energy $E$.

The Maslov canonical operator method, proposed in \cite{Maslov} (see also \cite{MF}),
  is a multidimensional version of the WKB method. In the case of the multi\-dimen\-sional Schr\"odinger operator
  $$
  \widehat{H} = \frac{\hbar^2}{2} \Delta + U(x)
  $$
  it is as follows.
  Let $M$ be a Riemannian manifold of dimension $d$
  with a metric $g_{jk}$ and $\Delta$ be the Laplace--Beltrami operator
  \eqref{lb}. Consider on the cotangent bundle $T^\ast M$ with the standard symplectic structure
  $$
  \Omega_0 = \sum_{k=1}^d dp_k \wedge dx^k
  $$
  a Hamiltonian system with the Hamiltonian function
  $$
  H(x,p) = \frac{1}{2}|p|^2 + U(x) = \frac{1}{2}g^{jk}(x)p_jp_k + U(x).
  $$
 Let $\Lambda$ be a Lagrangian submanifold in $T^\ast M$, invariant under the Hamiltonian flow and with an invariant measure $d\mu$ on $\Lambda$. Recall that a submanifold $T^\ast M$ is Lagrangian if it is $d$-dimensional and the restriction of the symplectic form onto it vanishes.

Let us take a covering $\{V_\alpha\}$ of the $\Lambda$ by simply connected open sets (charts) with local coordinates of the form   $x^{j_1}$, $\dots$, $x^{j_m}$, $p_{k_1}$, $\dots$, $p_{k_{d-m}}$. Let us take a partition of unity on $\Lambda$
 $\{\phi_\alpha\}$: $\phi_\alpha: \Lambda \to \RR, 0 \leq \phi_\alpha \leq 1, \sum \phi_\alpha \equiv 1$,
 $\phi_\alpha =0$ outside $V_\alpha$.

 The canonical operator acts on differentiable functions $u(s)$ on $\Lambda$ and
 maps them to functions on $M$.
 It is defined for each chart $V_\alpha$ with canonical coordinates $(x^I,p_I)$:
  $$
  K^\hbar_\Lambda(V_\alpha)(\phi_\alpha u)(x)
  $$
and the general operator is
\begin{equation}
\label{cansum}
\left(K^\hbar_\Lambda u\right)(x) = \sum_\alpha C_\alpha K^\hbar_\Lambda(V_\alpha)(\phi_\alpha u)(x),
\end{equation}
where $C_\alpha$ are some constants.

The operator $K^\hbar_\Lambda(V)$ has the simplest form in the case when a chart $V$ is projected onto a domain in $M$ with coordinates $x^1,\dots,x^d$:
\begin{equation}
\label{quasi2}
K^\hbar_\Lambda(V)(u)(x) = e^{\frac{i}{\hbar}S(y)} \sqrt{\frac{d\mu(y)}{\sqrt{ g}dx}} u(y),
\end{equation}
where $\pi: \Lambda \to M$ is the projection of $\Lambda$ onto $M$, $\pi(y)=x$ and $\sqrt{g}dx$ is the volume form on $M $ ($g = \det(g_{jk}), dx = dx^1\wedge \dots \wedge dx^d$).
Here
\begin{equation}
\label{action}
S(y) = \int_{y_0}^y \sum_{k=1}^d p_k dx^k
\end{equation}
is the action functional that is obtained by integrating along a path in $V_\alpha \subset \Lambda$ from 
some initial point $y_0$ to $y$. Because $V_\alpha$ is simply connected,  
$d(\sum p_kdx^k) = \Omega_0$ and $\Omega_0 \vert_\Lambda = 0$, the value of the integral does not depend on the choice of path.

Let us compare \eqref{quasi1} and \eqref{quasi2}. Let $\gamma$ be a periodic trajectory of a particle in 
a one-dimensional potential field. It is a Lagrangian submanifold in $T^\ast \RR$.
The invariant measure on it has the form
$$
d\mu = \frac{dx}{p} \ \ \ \mbox{and} \ \ \ \sqrt{\frac{d\mu}{dx}} = \frac{1}{\sqrt{p}} .
$$
A closed trajectory $\gamma$ with inflection points, at which $p=0$, removed is divided into 
two intervals, which are projected onto the interval of the line. 
They define two charts on $\gamma$, are which passed with respect to $x$ in different directions, and hence the signs of $\pm$ at $\int pdx$ arise.
The condition \eqref{bz} is replaced in a multidimensional situation by the following quantization condition:
for any closed curve $\gamma$ on $\Lambda$ we have
\begin{equation}
\label{qm}
\frac{1}{2\pi \hbar} \int_\gamma \sum_k p_k dx^k - \frac{\mu(\gamma)}{4} = n, \ \ n \in \ZZ,
\end{equation}
where $\mu(\gamma)$ is the Maslov index of the curve $\gamma$.
In the one-dimensional situation, \eqref{qm} reduces to \eqref{bz}.

Quasi-classical eigenfunctions of $\widehat{H}$ are constructed from Lagrangian sub\-ma\-ni\-folds lying at the energy level $E$ and satisfying the quantization condition \eqref{qm}. To construct them, the canonical operator is applied to the function $u \equiv 1$ on $\Lambda$:
$$
\psi(x) = (K^\hbar_\Lambda (1))(x), \ \ \ \widehat{H}\psi = E\psi \, \mathrm{mod}\, O(\hbar^2 ).
$$
In the one-dimensional case \eqref{quasi1} the analogue of $u$ is
$$
 e^{\sum_{k\geq 2} f_k(x) \left(\frac{\hbar}{i}\right)^{k-1}} = 1 \, \mathrm{mod}\, O( \hbar).
$$

\section{Quasi-classical quantization of magnetic monopoles}

Let us move on to extension of the multidimensional WKB method to the case of magnetic monopoles,
proposed in \cite{KT20}.

Let we have  a magnetic field (a closed $2$-form) on the Riemannian manifold $M$ $F^{(0)}$ and a particle with charge $q$. Let us define a $2$-form
$$
F = q F^{(0)}.
$$
If it satisfies the quantization condition
$$
\left[\frac{F}{2\pi}\right] \in H^2(M;\ZZ),
$$
then on $M$ there exists a $U(1)$-bundle $L$ with the first Chern class $c_1(L) =
\left[\frac{F}{2\pi}\right]$ and on the bundle we can define a connection $A=(A_k)$ such that
$$
F = dA.
$$

When this system is quantized, the momentum is replaced as follows
$$
p_k \to p_k - A_k, \ \ k=1,\dots,d.
$$
The magnetic Laplacian takes the form
$$
\Delta^L = -\frac{1}{\sqrt{g}}\left(\frac{\partial}{\partial x^j} - iA_j\right) \sqrt{g} g^{jk}
\left(\frac{\partial}{\partial x^k} - iA_k\right).
$$

The motion of a particle in a magnetic field is described by a 
Hamiltonian system on $T^\ast M$ with the Hamiltonian function $H(x,p) = \frac{1}{2}|p|^2$
with respect to the twisted symplectic structure \cite{Novikov}:
$$
\Omega = \sum_k d(p_k - A_k) \wedge dx^k =
\sum_k dp_k \wedge dx^k - \sum_{j<k} F_{jk} dx^j \wedge dx^k
= \Omega_0 - F.
$$

If the form $F$ is exact, then the canonical operator can be applied to the system, as done in
\S 5. It is just needed to consider the Lagrangian submanifolds with respect to the twisted symplectic structure and define the action functional as
\begin{equation}
\label{action2}
S = \int d^{-1}(\Omega) = \int \sum_k (p_k-A_k)dx^k.
\end{equation}

Suppose the form $F$ is not exact.
The parallel translation of sections of the bundle $L$ along the curve $\gamma$ is given by the equation
$$
\dot{\gamma}^k \left(\frac{\partial}{\partial x^k} \psi - iA_k\psi\right) = 0,
$$
which can be rewritten as
$$
\dot{\gamma}^k \left(\frac{\partial}{\partial x^k} \log \psi -iA_k\right) = 0.
$$
Therefore, the expression
$$
e^{\int iA_k dx^k},
$$
defines a section of $L$.  

In \S 4, when presenting the general quantization scheme, we assumed that  $\hbar=1$.
Here $\hbar$ is a small parameter and we have to consider expressions of the form
\begin{equation}
\label{sec}
e^{\frac{i}{\hbar} \int \sum_k (p_k-A_k)dx^k}.
\end{equation}
They can be given a clear geometric meaning only if the quantization conditions $\hbar$ are met:
\begin{equation}
\label{quanth}
\frac{1}{\hbar} \in \NN \ \ \ \mbox{or} \ \ \ \hbar= 1,\frac{1}{2},\frac{1}{3},\dots, \frac{1}{N},\dots,
\end{equation}
since for $\hbar = 1/N$ the expression \eqref{sec} describes the section of  $L^N$, i.e., the $N$-th tensor power of  $L$.

\vskip3mm

{\sc The scheme for quasi-classical quantization of a magnetic monopole} is as follows \cite{KT20}:

1) a choice of a Lagrangian (with respect to the twisted form $\Omega$)
submanifold $\Lambda \subset T^\ast M$, on which the Hamiltonian is equal to the constant: $H \equiv \frac{E}{2}$ (unlike the case of the Schr\"odinger operators considered earlier, we omit the factor $\frac{1}{ 2}$ before $\Delta$);

2) a choice of a covering $\Lambda$ by charts such that $L$ is trivial 
over their projections onto $M$;

3) for each chart $V$, take a restriction $A_V = A_{V,k} dx^k$ of the connection onto it;

4) then construct the canonical operator according to the usual scheme, replacing $p_k$ everywhere with $p_k-A_k$.
In particular, for a chart projecting onto a domain from $M$, the operator $K^\hbar_\Lambda(V)$ is constructed by the formula \eqref{quasi2} with $S$ replaced by \eqref{action2}. For other charts 
this will require some modifications, for which we refer to \cite{KT20}.

For $\hbar=1/N$ the additional quantization condition on $\Lambda$ must also be satisfied
(it depends on $N = \hbar^{-1}$):
for any closed curve $\gamma$ on $\Lambda$
\begin{equation}
\label{qq}
N\left(\int_\gamma \sum_k p_kdx^k + h_A(\gamma)\right) = \frac{\pi}{2}\mu(\gamma) \ \mathrm{mod}\ 2\pi\ZZ ,
\end{equation}
where $e^{ih_A(\gamma)}$ is the holonomy of the projection of $\gamma$ onto $M$ with respect to the connection of $A$ and
$\mu(\gamma)$ is the Maslov index of the curve $\gamma$.

If the quantization condition is satisfied, we can correctly construct the operator $K^{1/N}_\Lambda$ and
the resulting expression
$$
\psi(x) = K^{1/N}_\Lambda(1)(x)
$$
will be a section of $L^N$ satisfying the equation
$$
\hbar^2 \Delta^{L^N} \psi = E \psi \, \ \mathrm{mod}\, O(\hbar^2),
$$
which, taking $\hbar= \frac{1}{N}$ into account, takes the form
$$
\Delta^{L^N} \psi = N^2 E \, \psi \ \mathrm{mod}\, O(1).
$$

Note that the eigensections and their quasi-classical approximations are
sections of the same bundle.

The search for series of almost eigenvalues now reduces to finding Lagrangian manifolds that satisfy the quantization condition \eqref{qq}.
In \cite{KT20} this was done for the Dirac monopole.

The Hamiltonian system corresponding to the Dirac monopole on a two-dimensional sphere (see \S\S 3-4),
describes the motion of a particle with charge $q=Ze$ on $S^2$ in the external magnetic field $F^{(0)} =
\frac{1}{2}\sin \theta d\theta \wedge d\varphi$. This system is integrable and its first two integrals have the form
$$
E = \frac{1}{2}(p_\theta^2 + \frac{1}{\sin^2 \theta} p_\varphi^2), \ \ \
P = p_\theta - \frac{Z}{2} \cos \theta.
$$

For a fixed value $|Z|=N \geq 0$
the exact eigenvalues of the magnetic Laplacian are
$$
E_{N,j} = j(j+1) + \frac{N}{2}(2j+1), \ \ \ j=0,1,\dots,
$$
and have multiplicities
$$
\mu_{N,j} = 2j+1 +N.
$$
At the same time, the Lagrangian tori $\{E = \mathrm{const}, P = \mathrm{const}\}$ satisfying \eqref{qq} give
almost eigenvalues
$$
\hat{E}_{N,j} = j(j+1) + \frac{N}{2}(2j+1) + \frac{1}{4}
$$
with multiplicities
$$
\hat{\mu}_{N,j} = \mu_{N,j} = 2j+1+N,
$$
i.e., the approximate spectrum is shifted by a constant $\frac{1}{4}$ while preserving the multiplicity.

\section{Final remarks}

As we have already noted, almost eigenvalues are found from invariant tori that satisfy the quantization conditions \eqref{qq}. It is also not difficult to write out the eigensections corresponding to them.
Recently, explicit formulas for the case of the Dirac monopole were obtained by Yu.A. Kordyukov and
by \cite{KT23}.

In \cite{KM} a different approach to constructing almost eigenvalues is proposed, the asymptotic quantization. It associates symbols on an arbitrary symplectic manifold with operators that act on the so-called bundles of wave packets on this manifold. A discussion of the asymptotic quantization of the magnetic Laplacian is given in \cite{BNS}, where it is emphasized that bundles of wave packets are more complex objects than 
$U(1)$-bundles, but locally in the coordinates the actions of the asymptotically quantized Laplacian and the Laplacian itself almost coincide.
In \cite{BNS}, a series of asymptotic eigenvalues for magnetic monopoles on closed hyperbolic surfaces in constant magnetic fields are found. They are constructed by using Lagrangian tori that satisfy quantization conditions. Formulas for wave packets (``asymptotic eigenfunctions'') have not yet been written out.

The difference between asymptotic quantization and quasi-classical quantization is indicated by 
the absence in it of analogues of the quantization condition for Planck's constant \eqref{quanth} and 
the fact that the eigenfunctions and their asymptotic approximations are sections of different bundles.

\vskip5mm
The author thanks Yu.A. Kordyukov for useful discussions.

The work was performed according to the Government research assignment 
for IM SB RAS, project FWNF-2022-0004.


\begin{thebibliography}{00}

\bibitem{BNS}
J. Br\"uning, R. V. Nekrasov, A. I. Shafarevich: 
Quantization of periodic motions on compact surfaces of constant negative curvature in a magnetic field, 
Math. Notes, 81:1 (2007), 28--36.

\bibitem{Curie}
P. Curie:
Sur la possibilit\'e d'existence de la conductibilit\'e magn\'etique et du magn\'etisme libre,
Journal de Physique, 3e s\'erie, t.III, 1894, 415--416.

\bibitem{Darboux}
G. Darboux:
Probl\`eme de m\'ecanique, Bulletin des Sciences Math\'ematiques et Astronomiques, S\'erie 2, {\bf 2} (1878) no. 1, 433--436.

\bibitem{Dirac31}
P. A. M. Dirac:
 Quantised singularities in the electromagnetic field, Proc. Roy. Soc. London Ser. A, {\bf 133} (1931), 60--72.

\bibitem{Dirac78}
P. A. M. Dirac: 
The monopole concept, Internat. J. Theoret. Phys., {\bf 17}:4 (1978), 235--247.

\bibitem{KM}
M. V. Karasev, V. P. Maslov: 
Nonlinear Poisson Brackets. Geometry and Quantization., Transl. Math. Monogr., 119, Amer. Math. Soc., Providence, RI, 1993, xii+366 pp.

\bibitem{KT19}
Yu.A. Kordyukov, I.A. Taimanov, Trace formula for the magnetic Laplacian,
Russian Math. Surveys, 74:2 (2019), 325--361.

\bibitem{KT20}
Yu.A. Kordyukov, I.A. Taimanov: 
Quasi-classical approximation for magnetic monopoles
Russian Math. Surveys 75:6 (2020), 1067--1088.

\bibitem{KT21}
Y.A. Kordyukov, I.A. Taimanov:
Trace formula for the magnetic Laplacian on a compact hyperbolic surface,
Regul. Chaotic Dyn., {\bf 27}:4 (2022), 460--476.

\bibitem{KT23}
Yu.A. Kordyukov, I.A. Taimanov:
Quiasiclassical approximation of monopole harmonics.

\bibitem{Maslov}
V. P. Maslov: 
Th\'eorie des Perturbations et
M\'ethodes Asymptotiques, Etudes mathematiques, Dunod, Gauthier-Villars, Paris, 1972, xvi+384 pp.

\bibitem{MF}
Maslov, V. P.; Fedoriuk, M. V.:
Semiclassical approximation in quantum mechanics,
Contemp. Math., 5
dath. Phys. Appl. Math., 7
D. Reidel Publishing Co., Dordrecht-Boston, Mass., 1981, ix+301 pp.

\bibitem{Novikov}
S. P. Novikov: 
The Hamiltonian formalism and a many-valued analogue of Morse theory, Russian Math. Surveys, 37:5 (1982), 1--56.

\bibitem{Poincare}
H. Poincar\'e:
Remarques sur une exp\'erience de M. Birkeland, Compt. Rend. Acad. Sci., {\bf 123} (1896), 530--533.

\bibitem{Shnir}
Ya. M. Shnir:
Magnetic monopoles, Texts Monogr. Phys., Springer-Verlag, Berlin, 2005, xviii+532 pp.

\bibitem{Tamm}
Ig. Tamm: 
Die verallgemeinerten Kugelfunktionen und die Wellenfunktionen eines Elektrons im Felde eines Magnetpoles, Z. Phys., 1931, № 3--4, 141--150.

\bibitem{Thomson}
J.J. Thomson:
Cathode rays, Philosophical Magazine, {\bf 44} (1897), 293--316.

\bibitem{WY75}
T.T. Wu, C.N. Yang:
Concept of nonintegrable phase factors and global formulation of gauge fields,
 Phys. Rev. D, {\bf 12} (1975), 3845--3857.

\bibitem{WY}
T.T. Wu, C.N. Yang:
Dirac monopole without strings: monopole harmonics, Nuclear Phys. B, {\bf 107}:3 (1976), 365--380.


\end{thebibliography}
\end{document}